\documentclass[12pt]{iopart}
\usepackage{setstack}
\usepackage{iopams}
\usepackage{epsf}
\begin{document}
\title[Critical behavior of the piezoresistive response]
{Critical behavior of the piezoresistive response in RuO$_2$-glass
composites.} 
\author{C. Grimaldi$^1$\footnote[3]{claudio.grimaldi@epfl.ch}, 
T. Maeder$^{1,2}$, P. Ryser$^1$, and S. Str\"assler$^{1,2}$} 
\address{$^1$ Institut de Production et Robotique, LPM,
Ecole Polytechnique F\'ed\'erale de Lausanne,
CH-1015 Lausanne, Switzerland}
\address{$^2$ Sensile Technologies SA, PSE, CH-1015 Lausanne, Switzerland}

\begin{abstract}
We re-analyse earlier measurements of resistance $R$ and piezoresistance $K$ in
RuO$_2$-based thick-film resistors. The percolating nature of transport in these 
systems is well accounted by values of the transport exponent $t$ larger than
its universal value $t\simeq 2.0$. Furthermore, we show that the RuO$_2$ volume
fraction dependence of the piezoresistance
data fit well with a logarithmically divergence at the percolation thresold.
We argue that the universality breakdown and divergent
piezoresistive response could be understood in the framework of a
tunneling-percolating model proposed a few years ago to apply in 
carbon-black--polymer composites.
We propose a new tunneling-percolating theory based on the segregated microstructure
common to many thick-film resistors, and show that this model 
can in principle describe the observed 
universality breakdown and the divergent piezoresistance.
\end{abstract}


\section{Introduction}
\label{intro}
The term thick-film resistors (TFRs) commonly indicates glass-conductor
composites based on RuO$_2$ (or also Bi$_2$Ru$_2$O$_7$, Pb$_2$Ru$_2$O$_6$, and IrO$_2$)
grains mixed and fired with glass powders of usually lead borosilicate type.
The most successful commercial applications of TFRs exploit their quite large
change in bulk resistivity when subjected to applied deformation. 
Given their low temperature sensitivity and high stability, TFRs
are therefore often used in sensor applications like pressure 
or force sensors \cite{prude,white}.

Besides their applications, TFRs are also a quite unique type of
metal-insulator composite due to their complex microstructure \cite{pike,kusy1,kubovy}, 
nonuniversal behavior of transport \cite{pike,kusy}, coexistence of quantum tunneling 
with percolating behavior \cite{pikeseager},
marked microscopic elastic heterogeneities \cite{grima1}. These peculiarities, which to our best
knowledge are only partially shared by other compounds, contribute to make the TFRs
an interesting playground for basic research on metal-insulator composites.

Concerning their microstructure, TFRs are often in a segregated structure regime
in which large regions of glass of about $1$$\mu$m in size constraint the much smaller
conducting grains (from $10$ $\mu$m to about $300$ $\mu$m) 
to be segregated in between the interstices of neighbouring glass grains. 
In addition, as a result of the firing process, there is strong evidence of diffussion
of nanosized particles of metal insider the glass, leading therefore to a modified
glassy region around the original metallic grains \cite{mene}. 

The role played by these ultrasmall conducting particles in transport has not
yet been elucidated. The observation that the original larger metallic grains
are separated by thin glass films of only $1$-$2$ nanometer thickness 
has suggested direct tunneling as the main mechanism of transport \cite{chiang}.
However, recent electric force microscopy measurements have revealed
that the diffused metallic ultrasmall particles can contribute in a substantial way
to the electrical connectivity of the conducting regions, suggesting therefore
a possible active role in transport of such a modified glass \cite{ale1,ale2}.

In spite of their differences, the above scenarios both indicate or are at least
consistent with tunneling transport (between adjacent original metallic grains or 
assisted by the small particles diffused in the glass). 
Furthermore, the tunneling hypothesis is sustained also by the high 
values of the piezoresistive response, {\it i. e.} the sensitivity of 
electrical resistance upon applied strain, and the variable-range-hopping like
temperature dependence of transport at low temperatures \cite{temp}.

The above picture, together with the percolating behavior observed as a function 
of the metallic volume concentration,
suggestes that TFRs can be regarded as a sort of segregated tunneling-percolation 
systems: {\it i. e.} a non-homogeneous mixture of metal-insulating phases
in which tunneling cohexists with percolation. A typical homogeneous counterpart
is encountered
in some carbon-black--polymer composites where the tunneling-percolating
behavior takes place within a homogeneous and random arrangement of conducting and 
insulating phases \cite{balb1}. 

Among the different unusual aspects of TFRs, the often observed nonuniversality
of transport is certainly one of the most interesting ones, both from the point of view
of material science and basic statistical physics. According to the classical
theory of transport in percolating systems, the resistance $R$ of a metal-insulator
composite with metallic volume concentration $x$ follows a power-law behavior
of the form \cite{kirk,stauffer}:
\begin{equation}
\label{resis}
R\simeq R_0(x-x_c)^{-t},
\end{equation}
where $R_0$ is a prefactor, $x_c$ is the percolation critical concentration
and $t$ is the transport critical exponent. Random resistors network theories
predict that $R_0$ and $x_c$ depend on the microscopic details of the composite, 
such as the arrangement of the conducting phase within the sample and the
elemental resistances connecting two neighbouring sites \cite{kirk,stauffer}. 
Conversely, the value of the transport exponent $t$ is quasi-universal: for
percolation on a lattice of resistances with non-pathological distributions, 
$t$ depends only upon lattice dimensionality $D$ \cite{stauffer}. According
to the most recent and accurate calculations, $t\simeq 2.0$ for $D=3$
\cite{batrouni,clerc}.

Deviations from universality have been repeatedly reported for various composites
\cite{balb1,rubin} and, notably, for RuO$_2$-based TFRs which can display values of $t$ of
about $t\simeq3$-$4$ \cite{kusy,carcia} or even up to $t\simeq 7.0$ \cite{pike}. 
It is the subject of this paper to investigate further on 
the universality breakdown of TFRs and its microscopic origin by re-analysing
earlier data on transport and piezoresistance. We show that reported
metallic concentration dependences of the piezoresistive response for different
RuO$_2$-based TFRs imply tunneling-distance dependence of the exponent $t$. 
We argue that a possible origin
of such universality breakdown stems from tunneling-percolating processes
with strong fluctuations in tunneling distances, as previously proposed by
Balberg for $D=3$ homogeneous random arrangements of conducting and insulating 
phases \cite{balb2}. Finally, we propose a model of TFRs where segregated structures
and fluctuating tunneling distances are treated on the same level.

\section{Transport and piezoresistance of RuO$_2$-glass composites}
\label{transport}
Among the several published experimental results concerning the transport
properties of TFRs, many of them report the contemporary measurement of
resistance and piezoresistance on commercial samples for which, however, the 
actual compositions are not known with sufficient accuracy. Since we are interested
on the evolution of transport as the conducting concentration is varies in a controlled way,
the literature based on commercial TFRs is of little use.
However, we have been able to single out two published sets of data of RuO$_2$-based TFRs 
\cite{carcia,tamborin}
which, to the best of our knowledge, are the only reports providing both
the sheet resistance $R_\Box$ and the piezoresistance as a function
of the RuO$_2$ volume concentration $x$. 
According to Refs.\cite{carcia,tamborin},
both sets of measurements have been done
for RuO$_2$ powders with different specific surface areas (SA) mixed with
borosilicate \cite{carcia} or high lead silicate \cite{tamborin} glasses,
and the final resistors were obtained after standard firing cycles.

\begin{figure}[t]
\begin{center}
\epsfxsize=20pc
\epsfbox{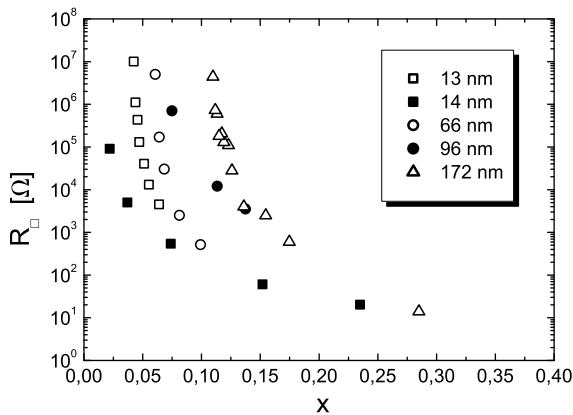}
\end{center}
\caption{Sheet resistance $R_\Box$ as a function of RuO$_2$ volume
fraction $x$ for different grain sizes of the metallic powders.
Open symbols: Ref.\cite{carcia}; full symbols: Ref.\cite{tamborin}}
\label{fig1}
\end{figure}

\begin{figure}[t]
\begin{center}
\epsfxsize=20pc
\epsfbox{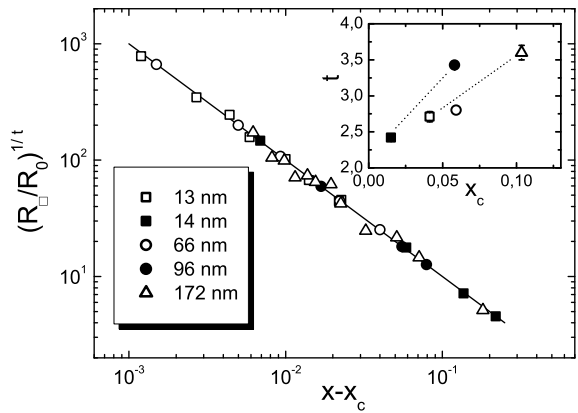}
\end{center}
\caption{Universal log-log plot relating $(R_\Box/R_0)^{1/t}$ with the 
distance $x-x_c$ from the critical concentration.
Open symbols: Ref.\cite{carcia}; full symbols: Ref.\cite{tamborin}.
The straight line is equation (\ref{resis}). Inset: $t$ {\it vs} $x_c$ obtained by the
fitting of equation (\ref{resis}) to the experimental data. The dotted lines are a 
guide to the eye.}
\label{fig2}
\end{figure}

In figure (\ref{fig1}) we report the sheet resistance $R_\Box$ data as a function
of RuO$_2$ volume fraction for different RuO$_2$ grain sizes $\Phi$ as extracted from
the SA values \cite{noteSA}.
The open symbols refer
to the data reported in Ref.\cite{carcia}, while the full symbols 
to Ref.\cite{tamborin}.
All sets of data show a strong increase of $R_\Box$ as $x$ decreases, reflecting
the vicinity to a percolation thresold where $R_\Box$ diverges.
The percolating nature of transport in these composites is demonstrated in the log-log
plot of figure (\ref{fig2}) where all data collapse in a single straight line
in accord with the power-law behavior of equation (\ref{resis}).
The values of $R_0$, the critical concentration $x_c$ and of the exponent $t$
have been obtained by least-square fits to (\ref{resis}), and are reported in 
table \ref{table1}.

There are a few points worth to be discussed. First, as it is clear from
the data of Ref.\cite{carcia} in table \ref{table1}, the critical concentration $x_c$ decreases
as the surface areas SA (or mean grain size $\Phi$) of the RuO$_2$ powders 
increases (decreases). This trend is also observed in the data of Ref.\cite{tamborin}
and it can be understood by realizing that finer RuO$_2$ powders occupy more likely
the interstitial regions between neighbouring glass grains. As shown by 
Refs.\cite{pike,kusy1,kubovy}, this enhanced segregation effect is reflected 
in a reduction of $x_c$. Note that, however, the SA {\it vs} $x_c$ 
data of Ref.\cite{carcia} do not 
merge with those of Ref.\cite{tamborin}. Maybe the different nature of the glass used in 
these two types of TFRs is at the origin of this discrepancy.

A second important point is that the values of the transport exponent $t$ are well
above the universal value $t\simeq 2.0$ of three dimensional lattices. This result
confirms other similar findings \cite{pike,kusy} and clearly
indicates  breakdown of transport universality in RuO$_2$-based TFRs. 
It is worth to remind that transport nonuniversality implies that the exponent $t$
acquires a dependence on microscopic details such as local intergrain resistances
and microstructure. In this respect, we note that the samples of Refs.\cite{carcia,tamborin}
display an increase of $t$ with the critical volume fraction $x_c$
as shown in the inset of figure (\ref{fig2}). We have noticed a qualitatively
similar behavior also in the transport measurements of Ref.\cite{szpytma},
where $t$ increases from $t=2.49$ to $t=5.38$ as $x_c$ is varied from
$x_c=0.0237$ to $x_c=0.0401$. Whether the increase of $t$ with $x_c$ is a general
feature of RuO$_2$-glass compounds is still an open question and further studies
are needed to settle this point. However it should be stressed that other ruthenate TFRs
(Bi$_2$Ru$_2$O$_7$ and Pb$_2$Ru$_2$O$_6$) display a quite different behaviour, {\it i.e.},
$t$ remains constant and very close to its universal value (Bi$_2$Ru$_2$O$_7$)
or slightly decreases as $x_c$ increases (Pb$_2$Ru$_2$O$_6$) \cite{carcia2}.

\begin{table}[t]
\caption{Specific surface areas (SA) and corresponding RuO$_2$ grain sizes $\Phi$
for the samples used in Refs.\cite{carcia,tamborin}.
$R_0$, $x_c$, and $t$ are the values of the parameters of equation (\ref{resis})
extracted by least square fits of the data reported in Refs.\cite{carcia,tamborin}.
$K_0$ and $B$ values are the corresponding fits of the piezoresistance data
to equation (\ref{gamma2}). The bracketed values are only indicative since they have
been obtained by fitting a set of only two points.}
\begin{indented}
\item[]\begin{tabular}{@{}lllllll}
\br
SA [m$^2/$g] & $\Phi$ [nm] & $R_0$ [$\Omega$] & $x_c$ & $t$ & $K_0$ & $B$ \\
\mr
$67^{\rm a}$ & $13$ & $0.145$ & $0.0413$ & $2.71\pm 0.07$  & $-13\pm 4$ & $3.8\pm 0.7$ \\
$13^{\rm a}$ & $66$ & $0.06$ & $0.0592$ & $2.80\pm 0.01$ & $-11\pm 3$ & $3.8\pm 0.6$  \\
$5^{\rm a}$ & $172$ & $0.0392$ & $0.1035$ & $3.6\pm 0.1$ & $-5\pm 2$ & $3.4\pm 0.5$  \\
$60^{\rm b}$ & $14$ & $0.515$ & $0.0151$ & $2.42\pm 0.03$ & $(-4.12)$ & $(1.73)$  \\
$9^{\rm b}$ & $96$ & $0.597$ & $0.0581$ & $3.4246$  & $-12\pm 1$ & $7.3\pm 0.4$  \\
\br
\end{tabular}
\item[] $^{\rm a}$ Ref.\cite{carcia}
\item[] $^{\rm b}$ Ref.\cite{tamborin}
\end{indented}
\label{table1}
\end{table}

\begin{figure}[t]
\begin{center}
\epsfxsize=20pc
\epsfbox{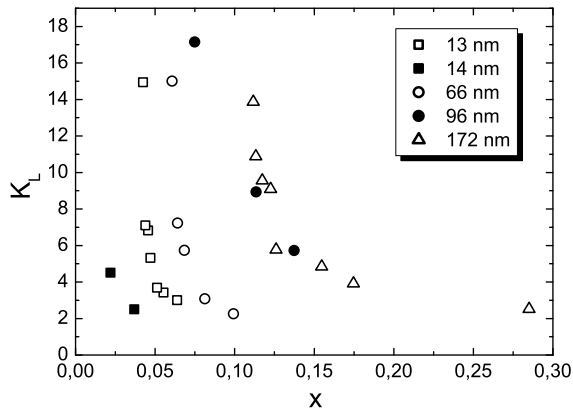}
\end{center}
\caption{Values of the longitudinal piezoresistance $K_{\rm L}$ as a function
of RuO$_2$ volume fraction $x$ for different specific areas of the metallic powders.
Open symbols: Ref.\cite{carcia}; full symbols: Ref.\cite{tamborin}.}
\label{fig3}
\end{figure}

The authors of Refs.\cite{carcia,tamborin} have also measured the RuO$_2$
concentration dependence of the longitudinal piezoresistance coefficient
$K_{\rm L}$ defined as:
\begin{equation}
\label{KL}
K_{\rm L}= \frac{d\ln(R)}{d\varepsilon}.
\end{equation}
The above quantity is also often called the (longitudinal) gauge factor and
it is obtained by recording the change of the sheet resistance caused
by a strain $\varepsilon$ applied along the direction of the external voltage drop.
When $\varepsilon$ is orthogonal to the voltage drop, the resistance change gives
rise to the transverse gauge factor $K_{\rm T}$. 
Both $K_{\rm L}$ and $K_{\rm T}$ depend on the geometry of the sample, but
they can be expressed in terms of the intrinsic longitudinal
and transverse piezoresistive coefficients, $\Gamma_\parallel$ and
$\Gamma_\perp$, defined as the logarithmic change of the {\it resistivity} under
uniaxial strains applied parallel and orthogonal to the voltage drop, 
respectively \cite{grima3}.
The values of $K_{\rm L}$ as
a function of $x$ are reported in figure (\ref{fig3}) for the same samples
of figures (\ref{fig1}) and (\ref{fig2}) \cite{noteprude}. Clearly, $K_{\rm L}$ seems to diverge 
as $x$ approaches to the same critical volume fraction values for which
$R_\Box$, figure (\ref{fig1}), goes to infinity.   
The origin of such a divergence has been previously discussed by
the authors of Ref.\cite{carcia}. Here, we disprove their arguments and
propose an alternative interpretation based on the universality breakdown
displayed by the data of figure (\ref{fig2}) and table \ref{table1}.

Our arguments go as follows. First, since for all the $x$ values considered
the samples are well within the critical region $|x-x_c|\ll 1$, then there is not marked
difference between longitudinal and transverse piezoresistive responses, and
the samples can be considered, in a first approximation, as electrically 
isotropic even if an uniaxial strain has been imposed to evaluate $K_{\rm L}$.
This is due to the tortuosity the current path has in flowing through the sample
when $x$ is close to $x_c$. More specifically, the piezoresistive anisotropy factor
defined as $\chi=(\Gamma_\parallel-\Gamma_\perp)/\Gamma_\parallel$ goes 
to zero as $(x-x_c)^\lambda$ with
$\lambda\simeq 0.5$ for three dimensional systems \cite{grima3}, so that for
$|x-x_c|\ll 1$ and apart from geometric factors, the $x$-dependence of uniaxial strain is 
basically indistinguishable from that obtained by applying equal strain along all directions. 
Hence, in the critical region,
it is a good approximation to evaluate $K_{\rm L}$ directly by differentiation
of equation (\ref{resis}) with respect to $\varepsilon$, just as done in
Ref.\cite{carcia}:
\begin{equation}
\label{gamma1}
K_{\rm L}\simeq \frac{d \ln (R_0)}{d\varepsilon}-
\frac{d}{d\varepsilon}\left[t\ln (x-x_c)\right].
\end{equation}
The first term is the usual contribution to the piezoresistance and it generally measures
the strain sensitivity of the mean junction resistance between two neighbouring
conducting particles.
Since, by construction, this term is independent of the RuO$_2$ concentration,
all the $x$ dependence of $K_{\rm L}$ reported in figure (\ref{fig3}) must
come from the last term of equation (\ref{gamma1}). The authors of Ref.\cite{carcia}
have assumed that $t$ and $x_c$ are not strain sensitive, and argued instead
that $x$, being a volume concentration, depends on $\varepsilon$. 
In this way, from equation (\ref{gamma1}), they obtained a $(x-x_c)^{-1}$ divergence
of $K_{\rm L}$ for $x\rightarrow x_c$ which fitted in a satisfactory way their
data [open symbols in figure (\ref{fig3})].
There is however a misconception in this reasoning. 
In fact, in percolation theory, $x$ is actually a 
measure of the concentration $p$ of (tunneling) intergrain junctions with finite
resistances present in the sample. 
Current can flow from one end to another
of the composite as long as a macroscopic cluster of junctions spans the 
entire sample. Hence the application of a strain, which is typically only of
order $\varepsilon\sim 10^{-4}$, can eventually change the value of the
junction resistances (affecting therefore $R_0$) but cannot modify the
concentration $p$ of junctions \cite{notedensity}. 

\begin{figure}
\begin{center}
\epsfxsize=20pc
\epsfbox{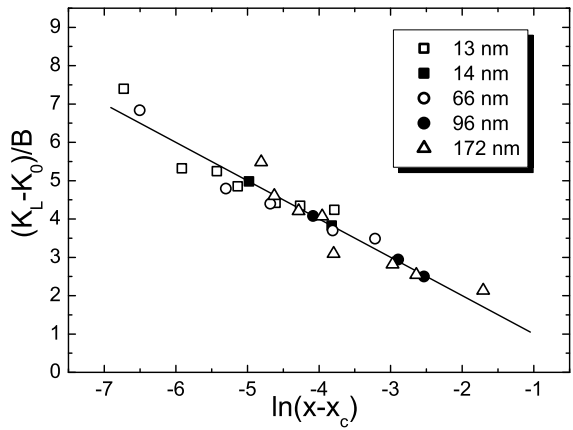}
\end{center}
\caption{Universal semi-log plot of $(K_{\rm l}-K_0)/B$. The straight line 
is equation (\ref{gamma2}).}
\label{fig4}
\end{figure}

The dependence of $x$ upon strain claimed in Ref.\cite{carcia} is therefore
erroneous and the $x$ dependence of $K_{\rm L}$ must have a different origin.
Since also $x_c$ is $\varepsilon$-independent, the only quantity left which
could have a strain dependence is the exponent $t$. This hypothesis is perfectly
in agreement with the universality breakdown demonstrated by the $t$
data of table \ref{table1}.
Hence, by allowing a nonzero derivative of $t$ with respect to $\varepsilon$
we obtain from equation (\ref{gamma1}):
\begin{equation}
\label{gamma2}
K_{\rm L}\simeq K_0+B\ln\left(\frac{1}{x-x_c}\right),
\end{equation}
where we have defined 
\begin{equation}
\label{gamma2b}
K_0=\frac{d\ln(R_0)}{d\varepsilon}, \,\,\,\,
B=\frac{dt}{d\varepsilon}.
\end{equation}
In figure (\ref{fig4}) we
report the results of fitting the data of figure (\ref{fig3})
to equation (\ref{gamma2}), represented by the straight line. 
The agreement is fairly good and indicates that the $K_{\rm L}$ data are indeed
consistent with a logarithmic divergence at $x\rightarrow x_c$. 
The fitted values of the parameters $K_0$ and $B$ are reported in 
the last two columns of table \ref{table1} where the main feature is that $B$ is always
positive while $K_0$ is always negative. This latter result is quite interesting.
In fact, naively, one expects that if $R_0$ is a measure of the mean interparticle
tunnel junction resistance, then it is a good approximation 
to set $R_0\propto\exp({2a/\xi})$, where $a$ is the mean
tunneling distance and $\xi\propto 1/\square{V}$ is the tunneling factor (or localization length)
and $V$ is the tunneling barrier potential.
Under a tensile strain $\varepsilon$, the mean tunneling distance is enhanced
by $a\varepsilon$, while $\xi$ remains constant. In this situation therefore
$K_0\sim 2a/\xi$ is always positive, contrary to the $K_0$ values of 
table \ref{table1}.
We shall see in the next section that this apparent contradiction is solved by 
allowing for sufficiently strong fluctuations in the tunneling distances. 
In this case in fact the notion of mean tunneling distance $a$ can be
meaningless and $K_0$ can assume also negative values.

\section{Balberg's model}
\label{model}

In the previous section, the logarithmic divergence of $K_{\rm L}$ 
has been obtained by allowing a strain-dependence of the resistance
exponent $t$. From a microscopic point of view, this property is equivalent
to require that $t$ is a functional of the intergrain tunneling distances $r_{\rm tun}$
between two neighbouring RuO$_2$ particles.
In fact, an applied strain $\varepsilon$ would modify the
intergrain distance $r_{\rm tun}$ to $r_{\rm tun}(1+\varepsilon)$ leading 
therefore to a variation of $t$. 
Actually, a microscopic theory of this kind exists and has been
proposed by Balberg a few years ago as a possible mechanism of
nonuniversality in carbon-black--polymer composites \cite{balb2}. 
Briefly, the theory of Balberg goes as follows. 
Consider a random resistor network where the interparticle (bond)
conductance distribution is given by:
\begin{equation}
\label{distrig}
\rho(g)=p\, h(g)+(1-p)\delta(g),
\end{equation}
where $p$ is the fraction of bonds with finite conductance $g$ with
distribution $h(g)$. The resistor network is constructed in order to mimic
a system constituted by a random arrangement of spherical particles 
of given diameter $\Phi$ and it is assumed that electrons can
tunnel from one particle to its nearest neighbouring particle
with a tunneling junction conductance of the form:
\begin{equation}
\label{g}
g=g_0 e^{-2r_{\rm tun}/\xi},
\end{equation}
where $g_0$ is a constant, $\xi$ is the localization length assumed to be 
smaller than the average interparticle distance $a$, and $r_{\rm tun}=
r-\Phi$ is the distance between the surfaces of two neighbouring spheres whose
centers are separated by $r$.
Since the particle centers are distributed more or less randomly, 
the distance $r$ depends on the particular couple of particles considered. 
Hence to obtain the distribution function $h(g)$ of the tunneling conductances 
one needs the distribution function $P(r)$ of inter-particles distances. 
Balberg proposed that the relevant distribution function is of the form:
\begin{equation}
\label{balb1}
P(r)=\frac{e^{-r/a}}{a},
\end{equation}
where $a$ is the mean inter-sphere distance and $a\gg \Phi$. 
By using
\begin{equation}
\label{distri1}
h(g)=\int_\Phi^{\infty}\!dr\,P(r)\delta\!\left(g-g_0 e^{-2r_{\rm tun}/\xi}\right),
\end{equation}
then the distribution function of tunneling conductances reduces to 
\begin{equation}
\label{distri2}
h(g)=\frac{1-\alpha}{g_0}\left(\frac{g}{g_0}\right)^{-\alpha},
\end{equation}
where 
\begin{equation}
\label{alpha}
\alpha=1-\frac{\xi}{2a},
\end{equation} 
is supposed to vary between $\alpha=0$ and $\alpha=1$.
As it is well known, distributions
of the form of equation (\ref{distri2}) lead to nonuniversal behavior of the
transport exponent $t$ \cite{kogut}. 
In fact, according to renormalization group results and highly disordered
resistor networks analysis,
the overall resistance close to the percolation thresold is given by equation 
(\ref{resis}) with \cite{machta,ledoussal}:
\begin{equation}
\label{distri3}
t=\cases{ t_0 & if $ (d-2)\nu+1/(1-\alpha)< t_0 $\\
(d-2)\nu+1/(1-\alpha) &  if $(d-2)\nu+1/(1-\alpha)> t_0$ \\}
\end{equation}
where $t_0$ is the universal value of the exponent ($t_0\simeq 2.0$ in three 
dimensions), $d$ is the dimension of
the system, and $\nu$ is the correlation-length exponent ($\nu\simeq 0.88$ in
three dimensions). We have therefore arrived at the result that, as soon as $\alpha$
is sufficiently large, the transport exponent depends upon the tunneling distance
$a$ and so can be affected by an applied external strain $\varepsilon$.
Hence, from equations (\ref{resis}), (\ref{gamma1}), and (\ref{distri3}) it is easily
found that the piezoresistance $K_{\rm L}$ follows equation (\ref{gamma2}) with
\begin{equation}
\label{Bbalb}
B=\frac{dt}{d\varepsilon}=\frac{1}{1-\alpha}.
\end{equation} 
Since $\alpha < 1$, the parameter $B$ is always positive, in agreement therefore with
the sign of the $B$ values reported in table \ref{table1}. 

\begin{figure}[t]
\begin{center}
\epsfxsize=20pc
\epsfbox{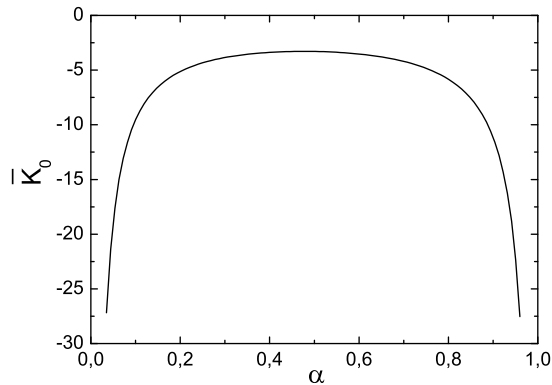}
\end{center}
\caption{$\bar{K}_0$ in the effective medium approximation [equation (\ref{ema3})]
as a function of the tunneling parameter $\alpha=1-\xi/2a$.}
\label{fig5}
\end{figure}

The distribution function of equation (\ref{distri2}) can also explain 
the negative values of $K_0$ shown in the last column
of table \ref{table1} and that, as we have pointed out in the last section, 
are not expected when ordinary interparticle
conductance distributions are considered. In fact, let us consider an
effective medium approximation (EMA) to the bond conductance distribution of
equations (\ref{distrig}) and (\ref{distri2}):
\begin{equation}
\label{ema1}
\int dg \rho(g)\frac{\bar{G}-g}{g+2\bar{G}}=0,
\end{equation}
where $\bar{G}$ is the effective macroscopic conductance \cite{kirk}. 
For $|p-p_c|\ll 1$
(where $p_c=1/3$ in EMA) the integral in equation  (\ref{ema1}) can be evaluated
by setting $x=g/\bar{G}$ with $\bar{G} \ll 1$ \cite{kogut}. 
The resulting resistance $\bar{R}=1/\bar{G}$ is then:
\begin{equation}
\label{ema2}
\bar{R}\simeq g_0^{-1}\left[\frac{3\sin(\pi\alpha)}{2^{1-\alpha}\pi(1-\alpha)}\right]
^{-1/(1-\alpha)}(p-p_c)^{-1/(1-\alpha)}.
\end{equation}
from which the EMA piezoresistance $\bar{K}$ reduces to $\bar{K}=\bar{K}_0+
1/(1-\alpha)\ln(p-p_c)^{-1}$
where:
\begin{equation}
\label{ema3}
\bar{K}_0=-\frac{1}{1-\alpha}\ln\left[\frac{3\sin(\pi\alpha)}{\pi (1-\alpha)}\right]
-\left[\frac{1}{1-\alpha}+\frac{\pi}{\tan(\pi\alpha)}\right].
\end{equation}
As shown in figure (\ref{fig5}), where the above expression is plotted
as a function of $\alpha$, $\bar{K}_0$ is always negative in agreement therefore
with the sign of the $K_0$ values extracted by the fits to the experimental data of
Refs.\cite{carcia,tamborin} (table \ref{table1}). 
The possibility of having $K_0 < 0$ stems from the fact that the distribution function 
(\ref{balb1}) has variance equal to the mean intergrain distance $a$. Hence the resistance prefactor 
$R_0$ is no longer a measure of the mean intergrain tunnel junction resistance so that
the arguments of the last section do not apply.

The above results for $B$ and $K_0$ have been derived by a straightforward derivative
of $R$ with respect to the applied strain $\varepsilon$. However, elastic heterogeneities
can have important effects. This should be particularly true for RuO$_2$-glass
systems where quite stiff metallic particles are embedded in a much softer glass.
The main effect of such elastic heterogeneity is that the external applied strain
induces highly varying local strains which depend on the relative bulk moduli
of the metallic and insulating phases and on the microstructure \cite{grima1}. We
can approximately describe the main effect of elastic heterogeneity by 
arguing that the tunneling distances are affected by a local 
strain $\varepsilon_{\rm loc}\simeq A\varepsilon$, where $A$ is a function of the 
relative bulk moduli of the conducting and insulating phases.  In this way, 
$B$ and $K_0$ of equation (\ref{gamma2b}) reduce to:
\begin{equation}
\label{gamma2c}
K_0=A\frac{d\ln(R_0)}{d\varepsilon_{\rm loc}}, \,\,\,\,
B=A\frac{dt}{d\varepsilon_{\rm loc}}.
\end{equation}
Since the local strains are concentrated mainly within the softer (insulating) phase through 
which the electrons tunnel, then $A$ is expected to be larger than the unity \cite{grima1}.
The elastic heterogeneity has therefore the main effect of amplifying the piezoresistive
response by enhancing the absolute values of $B$ and $K_0$, however leaving their signs unchanged.

\section{Tunneling and percolation in segregated compounds}
\label{disc}

In the previous sections we have provided evidences that RuO$_2$-based
thick-film resistors have nonuniversal behavior of transport and that
their piezoresistance response fits well with a logarithmic divergence about the
critical concentration $x_c$. Furthermore, we have shown that such
a divergent piezoresistance arises naturally from a specific model of
tunneling transport in percolative composites proposed some years ago by Balberg.
Although these results are quite encouraging, we are going to see that there
are several open questions
regarding Balberg's theory in general and its applicability to TFRs in particular.

\begin{figure}[t]
\begin{center}
\epsfxsize=20pc
\epsfbox{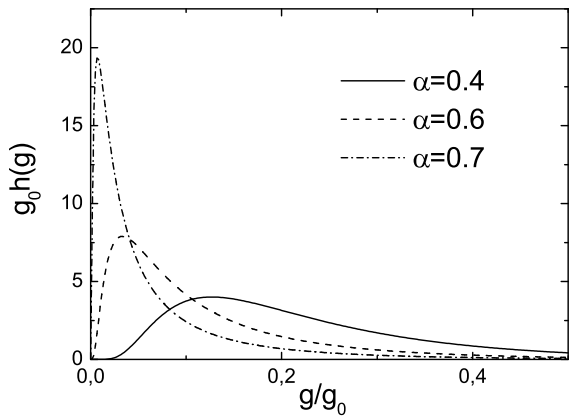}
\end{center}
\caption{tunneling conductance distribution function of equation (\ref{hertz2}) 
resulting from a random arrangement of conducting spheres in three dimensional space.}
\label{fig6}
\end{figure}

Let us start by considering the interparticle distribution function
of equation (\ref{balb1}). The reasons which prompted Balberg to use
equation (\ref{balb1}) are discussed in his original paper \cite{balb2}, where
the interested reader can directly refer. However, if we literally consider a three
dimensional random arrangement of spheres, then instead of equation (\ref{balb1}) the
inter-sphere distance distribution function for low volume densities is more correctly given 
by the Hertz distribution \cite{hertz}:
\begin{equation}
\label{hertz}
P(r)=3\Gamma(4/3)^3\frac{r^2}{a^3}e^{-\Gamma(4/3)^3(r/a)^3},
\end{equation}
where $\Gamma(4/3)\simeq 0.893$. As shown in Ref.\cite{torqua}, the above expression
is also the exact distribution function for a random set of fully impenetrable spheres 
at arbitrary density.
For large $r$, equation (\ref{hertz}) falls
to zero more rapidly than equation (\ref{balb1}), leading to a quite drastic effect
on $h(g)$. In fact, from equations (\ref{g}), (\ref{distri1}) and (\ref{hertz}) 
and again assuming for simplicity $\Phi\ll a$, the resulting tunneling conductance
distribution function reduces to:
\begin{equation}
\label{hertz2}
h(g)=\frac{3\Gamma(4/3)^3(1-\alpha)^3}{g_0}\ln^2\left(\frac{g_0}{g}\right)
\left(\frac{g}{g_0}\right)^{1-\Gamma(4/3)^3(1-\alpha)^3\ln^2(g_0/g)}.
\end{equation}
The above expression is plotted in figure \ref{fig6} for different values
of $\alpha$. Clearly, in contrast to equation (\ref{distri2}), $h(g)$ does not
diverges for $g\rightarrow 0$ but instead it goes to zero at $g=0$. 
In Ref.\cite{torqua} very accurate analytical expressions for interparticle 
distance distribution functions of impenetrable (hard-core) spheres are reported.
We checked that $h(g=0)=0$ also for this case so that such a vanishing limit 
is a general feature of randomly distributed spheres in three dimensional space.
Since transport nonuniversality 
and the appearance of a tunneling distance dependence of the exponent $t$
have as necessary condition a divergent $h(g)$ behavior for $g\rightarrow 0$,
then the distribution function of conductances resulting from a random arrangement of
(impenetrable or not) spheres is not able to account
for nonuniversality and the consequent logarithmic divergence of the piezoresistive
response. Such a conclusion is quite at odds with Balberg's theory and its extension
for the piezoresistive problem discussed in the previous section. However there
are a couple of points worth to be stressed. First, as shown in figure \ref{fig6},
$h(g)$ has a maximum which shifts at lower values of $g$ and sharpens
as $\alpha$ is made larger. Already for $\alpha=0.6$, the peak of $h(g)$ occurs at 
$g/g_0\simeq 0.03 \ll 1$. In this situation, the critical region where
the resistance follows equation (\ref{resis}) with universal exponent 
$t=t_0\simeq 2.0$ is expected to be considerably narrowed. Hence if $p$
is not very close to $p_c$ the resistance could appear to have the form of
equation (\ref{resis}) with an apparent nonuniversal value of $t$.

\begin{figure}[t]
\begin{center}
\epsfxsize=20pc
\epsfbox{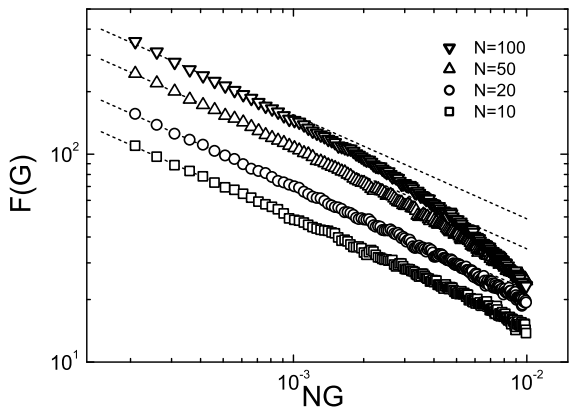}
\end{center}
\caption{Computer generated distribution functions $F(G)$ of the total conductances
$G$ of a series of $N$ tunnel junctions with conductances distributed according to
equation (\ref{distri2}) with $g_0=1$ and $\alpha=0.5$. The dotted curves
are $F(g)\propto G^{-0.5}$.}
\label{fig7}
\end{figure}

A second important point concerns the use of interparticle distance distribution
functions derived from random arrangements of spheres. 
In real composites, interactions between conducting and insulating phases and
microstructural inhomogeneities can easily make the idealized model of randomly
distributed spheres of little use. A remarkable example is provided actually
by thick-film resistors where the segregation effect due to 
the different sizes of glass and RuO$_2$ particles is completely missed by,
for example, equation (\ref{hertz}). However the segregation effect has
interesting consequences on the tunneling distance distributions. 
Let us consider the microstructure model of TFRs
proposed originally by Pike \cite{pike} (see also Refs.\cite{kusy1,kubovy}). In this model, the conducting
particles (approximated by spheres of diameter $\Phi$) are arranged 
in such a way to occupy narrow channels along the edges of large cubes 
of side $L\gg \Phi$ which represent the glass particles. 
These channels form a cubic lattice spanning the
whole sample and each channels can be occupied or free of conducting spheres.
We can imagine the channels to be quasi-one dimensional
so that each occupied  channel is actually given by a series of 
adjacent metallic particles separated by the sintered glass. 
Each channel can contain up to about
$L/\Phi$ number of spheres and, for typical TFRs, $L/\Phi$ is roughly between 
$50$ and $500$ depending on the relative sizes of the glass and conducting particles.
The interesting point here is that equation (\ref{balb1}) is the exact interparticle
distance distribution function for
a one-dimensional random arrangement of penetrable spheres. Hence, $h(g)$ given by
equation (\ref{distri2}) is the relevant conductance distribution function
for the junction conductances inside an occupied channel. The question is now
whether the total channel conductance $G$ has a power law diverging distribution
as well.
If a channel is occupied by $L/\Phi>N\gg 1$ spheres, then
the total channel conductance is that given by $N$ junctions in series
each with conductance $g_i$ ($i=1,\ldots ,N$):
\begin{equation}
\label{channel1}
G^{-1}=\sum_{i=1}^N\frac{1}{g_i}\simeq g_{\rm min}^{-1}+\frac{1-\alpha}{\alpha}
(N-1)g_{\rm min}^{-\alpha},
\end{equation}
where the distribution function (\ref{distri2}) has been used and $g_{\min}$ is the
minimum interparticle conductance for a row of $N$ particles.
In writing equation (\ref{channel1}),
we have set $g_0=1$ and have implicitly assumed that $N$ is large 
enough to replace the summation with an integral.
The quantity $g_{\rm min}$ (and so $G$) varies from channel to channel and
to find its distribution function $f(g_{\rm min})$ consider the probability that, 
in a given channel, $g_{\rm min}$ is larger than a given arbitrary value $\delta$:
\begin{equation}
\label{channel2}
{\rm Prob}(g_{\rm min}\ge\delta)=\int_\delta^1 \!dg_{\rm min}\, f(g_{\rm min}).
\end{equation}
The above probability must equal the probability that {\it none} of the $N$ junctions
within a channel has conductance smaller than $\delta$:
\begin{equation}
\label{channel3}
{\rm Prob}(g_{\rm min}\ge\delta)=\left[1-\int_0^\delta \!dg\, h(g)\right]^N=
\left(1-\delta^{1-\alpha}\right)^N.
\end{equation}
$f(g_{\rm min})$ is then readily found by equating the right hand sides of
equations (\ref{channel2}) and (\ref{channel3}) and by taking the derivative
with respect to $\delta$:
\begin{equation}
\label{channel4}
f(g_{\rm min})=(1-\alpha)N(1-g_{\rm min}^{1-\alpha})^{N-1}g_{\rm min}^{-\alpha}.
\end{equation}
The final distribution function $F(G)$ of channel conductances $G$ is then obtained by
combining equation (\ref{channel1}) with equation (\ref{channel4}). 
An asymptotic estimate of $F(G)$ for small $G$
can be derived from equation (\ref{channel1}) by realizing that
for $g_{\rm min}\ll (1/N)^{1-\alpha}$ the total channel conductance is dominated 
by the smallest junction conductance: $G\simeq g_{\rm min}$. 
Hence in this limit and from equation (\ref{channel4}): 
\begin{equation}
\label{channel5}
F(G)\propto G^{-\alpha},\,\,\,{\rm for}\,\,G \ll (1/N)^{1-\alpha}.
\end{equation}
The validity of the above limit is demonstrated in figure \ref{fig7}
where we have plotted computer generated values of $F(G)$ for $\alpha=0.5$
and for different number $N$ of junctions. Clearly, in accord
with equation (\ref{channel5}), the numerical distribution function diverges
as $G^{-\alpha}$ independently of $N$.

The set of channels which form the cubic lattice in our model
have therefore conductances $G$ with diverging distribution
at small $G$. In this situation therefore, by following the 
arguments of the  previous section, transport universality breaks down and
logarithmically divergent piezoresistive response sets in as long
as $\alpha$ is sufficiently large.
This interesting result can be made even more stringent by considering
the more realistic situation in which the spheres inside the channels are
impenetrable. Following Ref.\cite{torqua}, the exact interparticle distance
distribution function for this case is:
\begin{equation}
\label{channel7}
P(r)=\frac{1}{a-\Phi}e^{-(r-\Phi)/(a-\Phi)},
\end{equation}
where $r > \Phi$ and, again, $\Phi$ and $a$ ($a>\Phi$) are the diameter of the spheres
and the mean interparticle distance, respectively. By using equations (\ref{g})
and (\ref{distri1}) and by following the above analysis, 
it is straightforward to find that the distribution function
$h(g)$ within a channel and the channel conductance distribution $F(G)$
are again given by equations (\ref{distri2}) and (\ref{channel5}), respectively,
where however $\alpha$ of equation (\ref{alpha}) is now replaced by:
\begin{equation}
\label{channel8}
\tilde{\alpha}=1-\frac{\xi/2}{a-\Phi}.
\end{equation}
This result has not required the hypothesis that $r\gg\Phi$, 
as in the original formulation of Balberg, and has therefore a more general
validity. A more complete analysis of our model including Monte Carlo calculations 
and a generalization to spheres with distributed diameters will be
presented elsewhere. 

Before concluding this section, it is worth to mention that in 
modelling the microstructure of TFRs along the lines of Pike \cite{pike} we have
implicitly considered the metallic particles as more or less the original 
grains used in fabricating the TFRs. We have therefore not addressed the role
of the small metallic clusters of nanometer size which have been shown to occupy
the space between two neighbouring RuO$_2$ grains as result of metal-glass
interactions \cite{mene,ale1,ale2}. It could be argued that these particles effectively
reduce the tunneling potential barrier separating the metallic larger grains,
enhancing therefore $\xi$. Further investigations and possible modifications
of our model are therefore necessary for a more complete description
of transport in TFRs.

\section{Conclusions}
\label{concl}

In summary, we have confirmed that the sheet resistance of 
RuO$_2$-based thick-film resistors \cite{carcia,tamborin} 
follows equation (\ref{resis})
with transport exponent $t$ larger than the universal value $t\simeq 2.0$
characteristic of three dimensional systems \cite{batrouni,clerc}. Furthermore,
we have shown that the piezoresistance data taken on the same samples 
\cite{carcia,tamborin} are
consistent with a logarithmic divergence close to the percolation thresold.
We have argued that such divergence stems from a tunneling-distance dependence
of the transport exponent $t$ and an analysis of Balberg's 
theory \cite{balb1,balb2} of tunneling-induced
universality breakdown seems to qualitatively reproduce some of the experimental findings
of Refs.\cite{carcia,tamborin}. 
Furthermore, we have adapted Balberg's theory, which was originally proposed for
homogeneously random arrangements of conducting particles in an insulating host,
to the highly segregated structure of TFRs. We have shown that this model
naturally gives rise to a diverging conductance distribution function leading 
to universality breakdown of transport and logarithmically divergent piezoresistive
response at the percolation thresold.

\section{Acknowledgments}
We thank J. Dutta for stimulating and interesting discussions.
This work is part of TOPNANO 21 project n.5947.1.

\end{document}